\documentclass[twocolumn,showpacs,preprintnumbers,amsmath,amssymb,aps,prb,superscriptaddress,10pt]{revtex4-1}

\usepackage{graphicx}
\usepackage{bm}
\usepackage{color}

\newcommand{\cumu}[1]{\langle\!\langle #1 \rangle\!\rangle}

\begin{document}

\title{Nanotransformation and current fluctuations in exciton condensate junctions}
\author{H. Soller}
\affiliation{Institut f\"ur Theoretische Physik,
Ruprecht-Karls-Universit\"at Heidelberg,\\
 Philosophenweg 19, D-69120 Heidelberg, Germany}
\author{F. Dolcini}
\affiliation{Dipartimento di Scienza Applicata e Tecnologia, Politecnico di Torino, I-10129 Torino, Italy}
\author{A. Komnik}
\affiliation{Institut f\"ur Theoretische Physik,
Ruprecht-Karls-Universit\"at Heidelberg,\\
 Philosophenweg 19, D-69120 Heidelberg, Germany}
\date{\today}

\begin{abstract}
We analyze the nonlinear transport properties of a bilayer exciton condensate that is contacted by four metallic leads by calculating the full counting statistics of electron transport for arbitrary system parameters. Despite its formal similarity to a superconductor the transport properties of the exciton condensate turn out to be completely different. We recover the generic features of exciton condensates such as
counterpropagating currents driven by excitonic Andreev reflections and make predictions for nonlinear transconductance between the layers as well as for the current (cross)correlations and generalized Johnson-Nyquist relationships.
 Finally, we explore the possibility of connecting another mesoscopic system (in our case a quantum point contact) to the bottom layer of the exciton condensate and show how the excitonic Andreev reflections can be used for transforming voltage at the nanoscale.
\end{abstract}


\maketitle

Transport in electronic bilayer systems has recently received increasing attention due to the possibility of observing the formation of   quantum macroscopic order in these systems. Indeed, when an electron layer and a hole layer are separated by an insulating barrier that is sufficiently thick to prevent inter-layer tunneling but sufficiently thin to induce interlayer Coulomb interaction, an excitonic condensate (EC) is predicted to form.\cite{keldysh1,*eisenstein1,*snoke,*eisenstein2,PhysRevB.59.R7825,*PhysRevLett.85.820}  Such condensate is a macroscopic quantum coherent state, in which electrons in one layer are bound to move coherently with holes in the other layer.
These  predictions have been confirmed in several experiments performed on GaAs quantum wells separated by an AlGaAs barrier, both in the quantum Hall regime at total filling factor $\nu=1$, see [\onlinecite{PhysRevLett.68.1383,*PhysRevLett.80.1714,*PhysRevLett.84.5808,*PhysRevLett.93.036801,*PhysRevLett.93.036802,*1367-2630-10-4-045018,*PhysRevLett.106.236807}] and, more recently, also at zero magnetic field.~\cite{PhysRevLett.101.246801,*PhysRevLett.102.026804}\\
So far, most theoretical  studies on transport properties in EC were concentrated on the linear response regime,\cite{PhysRevB.59.R7825,*PhysRevLett.85.820,su2008} with a special focus on Coulomb drag configurations.\cite{2011arXiv1108.2298M} Other  recent  works have considered the case of EC contacted to superconducting electrodes,\cite{PhysRevLett.104.027004,2011arXiv1108.1533P} whereas current fluctuation properties have   only been addressed for systems where interlayer Coulomb interaction is present but is not strong enough to lead to condensation.~\cite{PhysRevLett.84.5383,*PhysRevB.65.085317}\\
A remarkable advance in the field of EC is expected to arise from graphene bilayers. Such ECs are predicted to exhibit substantially higher critical temperature than ordinary semiconductor  realizations,\cite{PhysRevB.78.121401,su2008,lozovik2008} due to the weaker screening and the higher electron and hole densities that can be achieved in graphene.  Quite recently, systems of two graphene layers separated by a thin insulating  boron nitride film have been realized,\cite{2011arXiv1107.0115P} and transport experiments in these systems may become a reality in the near future.\\
%Due to the peculiar electronic properties of graphene, this may open up the possibility to observe transport regimes and phenomena that have not been addressed so far. A more exhaustive description of transport properties in EC bilayers is thus needed. \\
In this paper we  derive the full counting statistics (FCS) of an EC bilayer, providing its complete low-frequency transport characteristics.\cite{levitov1} This enables us to investigate not only the nonlinear conductance, but also the current noise and the higher current cumulants.   To this purpose we shall adopt the model developed in [\onlinecite{PhysRevLett.104.027004}], and evaluate  the cumulant generating function (CGF) of charge transfer via the nonequilibrium Green's function technique.~\cite{nazarov,*kindermann}
%%%%%%%%%%%%%%%%%%%%%%%%%%%%%%%%%%%%%
\begin{figure}
\includegraphics[width=6cm]{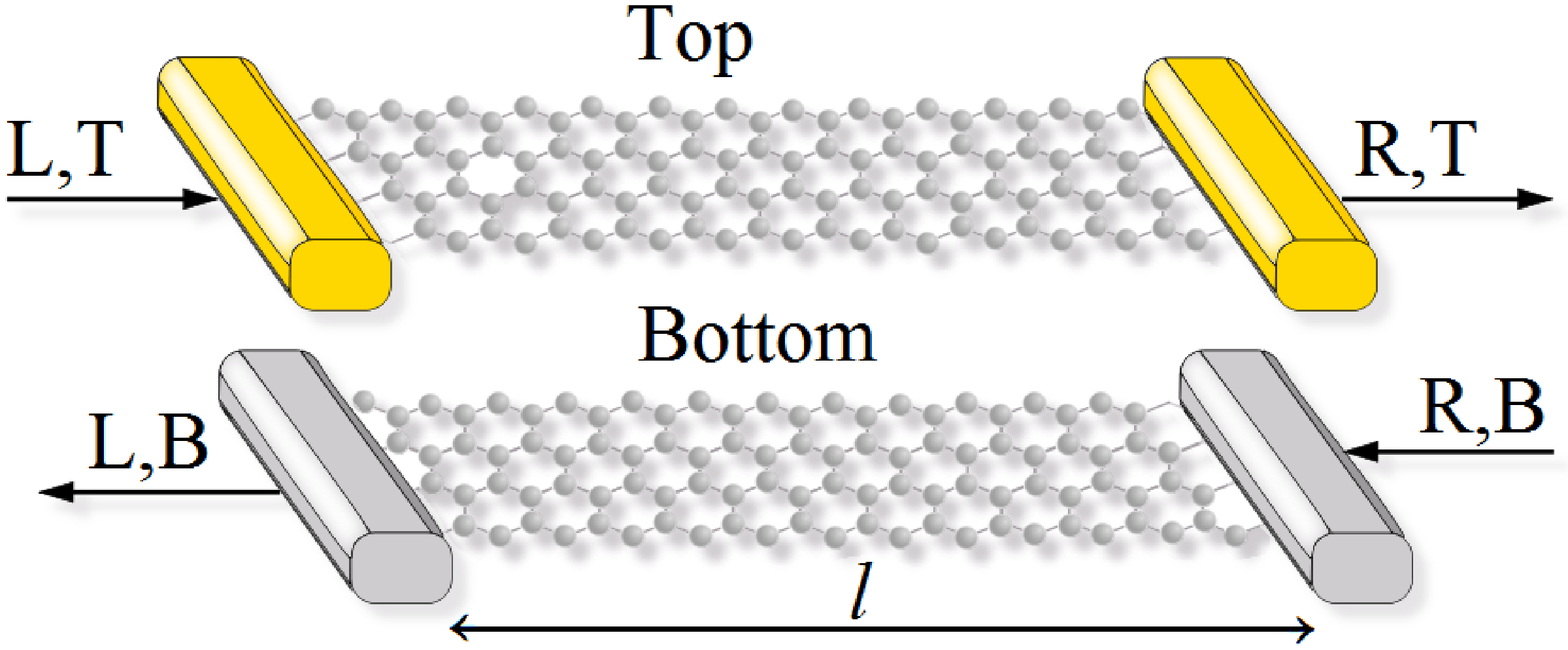}
\caption{(color online) Sketch of the experimental setup. The double layer EC is contacted with four metallic electrodes.}
\label{fig1}
\end{figure}
%%%%%%%%%%%%%%%%%%%%%%%%%%%%%%%%%%%%%
Moreover, we shall take a mesoscopic view on drag-counterflow geometries where the top layer is contacted by leads at different chemical potentials inducing a current in the bottom layer that is also part of another circuit.\cite{su2008} In our case we study a quantum point contact between the two leads of the bottom layer and explore the possibility of transforming current on the nanoscale.

The system, schematically depicted in Fig.~\ref{fig1}, consists of an electron-hole bilayer, where each layer is contacted to two metallic electrodes. While no inter-layer tunneling is assumed to occur, the two layers are coupled  via Coulomb interaction. The Hamiltonian modeling the system reads $H = H_n + H_T +H_{EC}$.
The term $H_n$ accounts for the four metallic electrodes, characterized by electrochemical potentials $\mu_{\alpha \sigma}$, Fermi distribution functions $n_{\alpha \sigma}$, and an energy-independent density of states $\rho_0$. $\alpha=L, R$ refers to the contacts on the left/right side of the bilayer, whereas $\sigma={T},{B}$ labels the top and bottom layer, respectively.
$H_T$ describes the particle tunneling between the layers of the EC and the metallic contacts
\begin{eqnarray}
H_T &=& \sum_{\sigma = {T}, \; {B}, \; \alpha = L, \; R} \gamma_{\alpha\sigma} (\alpha_\sigma^\dagger \Psi_\sigma + \Psi_\sigma^\dagger \alpha_\sigma)  \, , \label{HT}
\end{eqnarray}
where $\gamma_{L{T},  {B}}, \; \gamma_{R{T},  {B}}$ are the   tunneling amplitudes, $L_{T/B}, \; R_{{T}/{B}}$ the electron field operators for the four leads, and  $\Psi^{}_{{T},  {B}}$ the  field operators for electrons in the EC layers  at the position $x=0,l$ (for $L, R$), respectively. As spin is irrelevant in the effects we are investigating we consider a spinless system.
Finally $H_{EC}$ describes the EC bilayer. For the EC all important features we want to describe are captured by a simple one-dimensional model \cite{PhysRevLett.104.027004}
\begin{eqnarray}
H_{EC}= \int_{-l/2}^{l/2} \!\!\! dx  \, \,   \Psi^\dagger(x)  \left(\begin{array}{cc} H_{T} & \Delta  \\ \Delta^*  & H_{B} \end{array}\right) \Psi^{}(x) \label{Hex} \, ,
%H_{EC} &=& \sum_k \left(\epsilon_{k{T}} \Psi_{k{T}}^+ \Psi_{k{T}} + \epsilon_{k{B}} \Psi_{k{B}} \Psi_{k{B}}^+ \right) \nonumber\\
%&& + \Delta \sum_k \left(\Psi_{k{T}}^+ \Psi_{k{B}} + \Psi_{k{B}}^+ \Psi_{k{T}}\right), \label{Hex} \, ,
\end{eqnarray}
where $l$ is the longitudinal distance between the electrodes, $\Psi=(\Psi_{T},\Psi_{B})^T$ is the two-layer spinor, $H_{T}$ ($H_{B}$) describes the electron (hole) single-particle term of the top (bottom) layer.
The inter-layer Coulomb interaction is described~\cite{PhysRevB.78.121401} by an exciton order parameter $\Delta(x)$, which is  in general a space-dependent quantity. Its bulk absolute value $\Delta_0$ at equilibrium represents the excitonic gap and determines the excitonic correlation length  $\xi_{EC} = v_F/\Delta_0$. We use units such that $k_B = e = \hbar = 1$ and $G_0 = 2 e^2 /h$.

%The Hamiltonian~(\ref{Hex}) is reminiscent of the Bogolubov de Gennes  Hamiltonian for a superconductor, where the  layer indices ${T}$ and ${B}$  play the role of spin indices in BCS theory.\cite{PhysRev.108.1175} Formally, the system has thus some analogies with a hybrid normal-superconductor-normal (NSN) junction. There are, however, two important differences. First, while the BCS order parameter couples Cooper electron pairs, the EC order parameter couples electron-hole pairs, as it is clear from the last line of Eq. (\ref{Hex}), where the additional transformation of ${B}$ holes into electrons must be performed to recover the  BCS Hamiltonian.\cite{lozovik2} Second, while in a superconductor Cooper pairs belong to the same physical system, in a bilayer geometry electrons and holes live on separate  layers, which are characterized by different chemical potentials  and which are independently contacted.\cite{eisenstein2,10.1063/1.2132071} As a consequence, this geometry allows to selectively tune the electrochemical potential of each 'spin' species and to probe 'spin'-resolved transport.
%Because of these differences, the present bilayer setup is much richer than the NSN junction or   other normal-superconductor hybrid  structures  considered before.\cite{PhysRevLett.87.067006,*PhysRevB.50.3982,*2010EPJD..tmp..289S,*soller1}\\

The FCS is the probability distribution function $P({\bf Q})$ for the charges ${\bf Q} = (Q_{L{T}}, \; Q_{R{T}}, \; Q_{L{B}}, \; Q_{R{B}})$ to be transferred through the respective junctions during a (long) waiting time $\tau$, thereby allowing to compute not only  non-linear I-V, but also current noise and higher order cumulants. This information is encoded in the  CGF $\chi({\bm \lambda}) = \sum_{{\bf Q}} e^{i {\bf Q} {\bm \lambda}} P({\bf Q})$ where ${\bm \lambda} = (\lambda_{L{T}}, \; \lambda_{R{T}}, \; \lambda_{L{B}}, \; \lambda_{R{B}})$ are the measuring fields. The cumulants (irreducible moments) are then found from the respective derivatives of $\ln \chi({\bm \lambda})$.\cite{PhysRevLett.98.056603}
In order to obtain the CGF we adopt the approach of modifying the Hamiltonian by introducing a time-dependent counting field and relate $\chi({\bm \lambda})$ to the Keldysh Green's functions of the system.\cite{PhysRevB.73.195301} Such procedure allows for the calculation of the FCS for arbitrarily given parameters of (\ref{Hex}) and of the tunneling amplitudes in~(\ref{HT}).~\cite{PhysRevB.54.7366} The determination of the currents  and its cumulants in this hybrid structure, however, represents an essentially self-consistent problem, where the external currents depend on the electrochemical potentials of the two layers and on the excitonic order parameter, which in turn adjust to ensure charge conservation and no inter-layer tunneling, thereby affecting the external currents themselves. In order to proceed, some assumptions are thus necessary. In view of possible implementations with graphene, we shall consider a linear Dirac cone spectrum $H_0$ for the layers, oppositely shifted by e.~g. two external gates $\pm V_g$, so that $H_{T/B}=H_0\mp eV_{g}-\mu_{EC,T/B}$, where $\mu_{EC,T/B}$ are the electrochemical potentials.
It is sensible to focus  on the incoherent tunneling regime, $\xi_{EC}, l_{\phi} \ll l$, where $l_\phi$ is the dephasing length.
%\fab{\bf Do we really need this? The length $l$ (see Fig. \ref{fig1}) is also assumed sufficiently large to neglect electromagnetic coupling between the two junctions, i.e. mutual capacitance between the normal wires, propagating modes in the EC and similar effects as done for a superconducting wire in [\onlinecite{PhysRevB.82.134508}]}.
The condition $\xi_{EC} \ll l$ also implies that self-consistency effects on the space dependence of $|\Delta(x)|$ are negligible.~\cite{su2008,Gerd1981171} A space-dependent phase ${\rm arg}(\Delta(x)) \sim q x$, on the other hand, although essential to ensure that  the EC carries counterflowing currents in the bulk of the bilayer, is not necessary for evaluating the currents in the leads, which are of interest here.\footnote{In principle a finite $q$ also affects the gap. We neglect this effect here, as customary in similar situations concerning superconductors.} In contrast, self-consistency of the electrochemical potentials $\mu_{EC,{T}/{B}}$ of the two layers  is crucial to ensure current conservation in each layer~\cite{lambert,*PhysRevB.51.17999}
\begin{eqnarray}
\langle I_{L{T}} \rangle  =    \langle I_{R{T}}\rangle, \hspace{1cm} \langle I_{L{B}}\rangle = \langle I_{R{B}}\rangle. \label{self}
\end{eqnarray}
%Such condition also implies that no current is flowing between the top and bottom layer.\cite{PhysRevB.56.3296}
Under these assumptions, we have obtained the complete analytical expression for the CGF for all parameter regimes.
Such expression, which has been used for our numerical evaluation, is quite lengthy and we do not report it here. Nevertheless, all relevant ingredients of the CGF already appear in the limits of small bias ($\mu_{L,\sigma}, \mu_{R,\sigma} \ll \Delta$) and large bias  ($\mu_{L,\sigma}, \mu_{R,\sigma}  \gg \Delta$), where the expression of the CGF greatly simplifies,
%Indeed in these regimes the non interacting self-energy due to the EC is either real and purely off-diagonal or imaginary and purely diagonal, like in a superconductor, while the one due to the normal lead is always diagonal.
and acquires the following form on the left leads
\begin{widetext}
\small
\begin{eqnarray}
&& \left. \ln \chi \right|_{\lambda_{R \sigma =0}} = 2 \tau \int \frac{d\omega}{2\pi} \left( \sum_{\sigma = {T}, \; {B}} \ln \left\{1+ T_\sigma(\omega) \left[(e^{i \lambda_{L\sigma}} -1) n_{L\sigma} (1-f_{\sigma}) + (e^{-i \lambda_{L\sigma}} -1) f_{\sigma} (1-n_{L\sigma})\right]\right\} \theta\left(\frac{|\omega_\sigma| - \Delta}{\Delta}\right) \right. \nonumber\\
&& + \left. \ln \left\{1+ T_A(\omega) \left[(e^{i \lambda_{L{T}}} e^{-i \lambda_{L{B}}} -1) n_{L{T}} (1-n_{L{B}}) + (e^{i \lambda_{L{B}}} e^{-i \lambda_{L{T}}} -1) n_{L{B}} (1-n_{L{T}}) \right]\right\}\theta\left(\frac{\Delta - \max(|\omega_{T}|, \; |\omega_{B}|)}{\Delta}\right)\right), \label{cgf}
\end{eqnarray}
\end{widetext}
where the transmission coefficients are given by $T_\sigma(\omega) = 4\tilde{\Gamma}_{L\sigma} / (1+ \tilde{\Gamma}_{L\sigma})^2$ and $T_A(\omega) = 4\tilde{\Gamma}_A / (1+ \tilde{\Gamma}_A)^2$. The effective transparencies are parametrised by the EC density of states as $\tilde{\Gamma}_{L\sigma} = \Gamma_{L\sigma} |\omega_\sigma| / \sqrt{\omega_\sigma^2 - \Delta^2}$ and $\tilde{\Gamma}_A = \Gamma_{L{T}} \Gamma_{L{B}} \Delta^2 / [\sqrt{\Delta^2 - \omega_{T}^2} \sqrt{\Delta^2 - \omega_{B}^2}]$, where $\Gamma_{L\sigma} = \pi^2 \rho_{0L\sigma} \rho_{0E} \gamma_{L\sigma}^2/2$. The functions  $f_{{T}}$ and $f_{{B}}$ denote Fermi distributions for the quasiparticles in the separate layers and $\omega_{{T}, {B}} = \omega - \mu_{EC,T/B}$.
%The CGF for the right side has the same form and can be obtained by exchanging $\lambda_{L\sigma}$ by $\lambda_{R\sigma}$, $n_{L\sigma}$ by $n_{E\sigma}$, $n_{E\sigma}$ by $n_{R\sigma}$ and $L$ and $R$ as subscripts in the transmission coefficients.\\
\begin{figure}
\includegraphics[width=5.5cm]{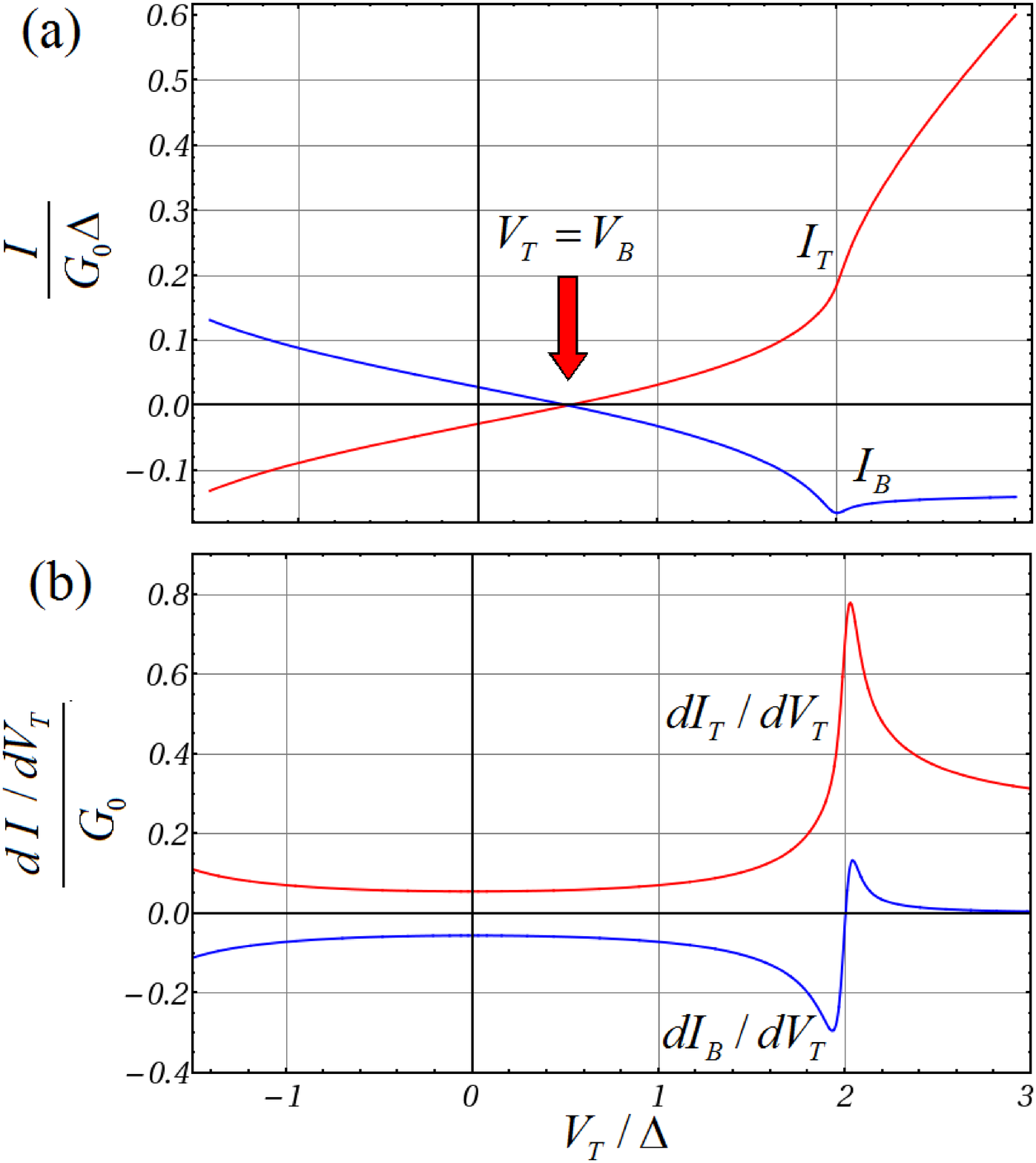}
\caption{(color online) (a) current in the top (red) and bottom (blue) layer as a function of $V_{T}$, for a fixed value $V_B=\Delta/2$. $\Gamma_{L{T}} = \Gamma_{L{B}} = 0.171$ corresponds to transmission of 0.5 for the uncoupled system, $T=0.01\Delta$. For $|V_T|,|V_B|<2 \Delta$ the bilayer exhibits counterpropagating currents, exciton blockade occurrs at $V_{B} = V_{T}$. (b) differential conductance $dI_T/dV_{T}$ (red) shows a resonance peak and tends to the typical value for a quantum point contact whereas the transconductance $dI_B/dV_{T}$ shows a resonance peak before vanishing at larger bias.}
	\label{fig2}
\end{figure}
The first line of Eq.~(\ref{cgf}) describes the supra-gap contribution, which is only due to single electron transport  and is characterized by the normal transmission coefficient $T_{\sigma}$. In contrast, the second line describes the sub-gap contribution due to the phenomenon of excitonic Andreev reflection~\cite{su2008}, consisting of an electron and a hole (traveling in {\it different} layers), which enter or leave coherently the bilayer in order for an excitonic pair to be transfered along the bulk of the system.\\
%These two basic processes directly affect the currents and their fluctuations, which are straightforwardly obtained from Eq.~(\ref{cgf}). \\
%Let us first discuss the currents.
The expression for the currents in the  left leads is
$
\langle I_{L \sigma} \rangle = - i \, \tau^{-1} \, \partial \ln \chi  /\partial  \lambda_{L\sigma}
$. Expressions for the rhs are obtained by  replacing $\lambda_{L{T}} \rightarrow - \lambda_{R{T}}, \; \lambda_{L{B}} \rightarrow - \lambda_{R{B}}$ in Eq.~(\ref{cgf}). Imposing the self-consistency condition (\ref{self})
determines $\mu_{EC\sigma}$, and one obtains  the final results for the two currents $I_\sigma \doteq \langle I_{L\sigma} \rangle = \langle I_{R\sigma}\rangle$. For simplicity, we consider the symmetric junction case, $\Gamma_{L{T}} = \Gamma_{R{T}}$ and $\Gamma_{L{B}} = \Gamma_{R{B}}$, and symmetrically applied biases $\mu_{L{T}} = - \mu_{R{T}} = - V_{T}/2$, $\mu_{L{B}} = - \mu_{R{B}} = -V_{B}/2$. In this case Eq.~(\ref{self}) is always fulfilled for $\mu_{EC{T}} = \mu_{EC{B}} = 0$.  \\
The  average currents are plotted in Fig. \ref{fig2}~(a) as a function of the top layer bias $V_T$, for a fixed value of the bottom layer bias $V_B$.
As one can see, because of the EC coupling, both $I_{T}$ and $I_{B}$ change, even when varying $V_T$ only.
In particular, for $|V_{T}|, |V_{B}| < 2\Delta$, one observes $
I_{T}  = - I_{B}$,   a signature that in the sub-gap regime  transport can only occur via  excitonic   counterpropagating currents in the bulk of the layers, which are transformed into electron and hole currents in the leads  through excitonic Andreev reflections.\cite{su2008}   Notice that for the value $V_T=V_B$ a current  locking occurs ($I_{T}  =   I_{B}=0$), because the EC cannot sustain currents driven by equally applied biases (exciton blockade). At $V_T= 2 \Delta$  excitonic pairs start to break up and the resulting electrons/holes get excited above the gap. This is clearly shown in  Fig.~\ref{fig2} (b), where the positive conductance exhibits a resonance peak, whereas the negative transconductance abruptly changes sign. At higher voltage values the EC plays a minor role, so that the conductance tends to the value of the case $\Delta=0$, and the transconductance vanishes, indicating that transport in the bottom layer is independent of the voltage applied to the top layer.

% For equal bias in the two layers the magnitude of the current becomes zero (exciton blockade), see also Fig. \ref{fig2} (a) for $(V_{B} - V_{T})/\Delta = 0$. If only one layer is biased an opposite current in the other layer is induced (drag), see Fig. \ref{fig2} (a) for $(V_{B}- V_{T}) / \Delta = -1$. If we further increase the bias difference between the layers the counterflowing currents increase (Fig. \ref{fig2} (a) for $(V_{B} -V_{T})/ \Delta = -2$). This resembles the transport features observable under phase bias instead of voltage bias.\cite{PhysRevLett.104.027004}  In Fig. \ref{fig2} (b) we also show the nonlinear conductance in the different situations. In the exciton blockade both conductances cross 0 whereas in the drag situation they are both zero as well but do not cross and increase upon further increase of the bias.\\

%%%%%%%%%%%%%%%%%%%%%%%%%%%%%%%%%%%%%%%%%%%%
%%%%%%%%%%%%%%%%%%%%%%%%%%%%%%%%%%%%%%%%%%%%
%%%%%%%%%%%%%%%%%%%%%%%%%%%%%%%%%%%%%%%%%%%%
%%%%%%%%%%     N O I S E   %%%%%%%%%%%%%%%%%
%%%%%%%%%%%%%%%%%%%%%%%%%%%%%%%%%%%%%%%%%%%%
%%%%%%%%%%%%%%%%%%%%%%%%%%%%%%%%%%%%%%%%%%%%
%%%%%%%%%%%%%%%%%%%%%%%%%%%%%%%%%%%%%%%%%%%%

\begin{figure}
\includegraphics[width=5.5cm]{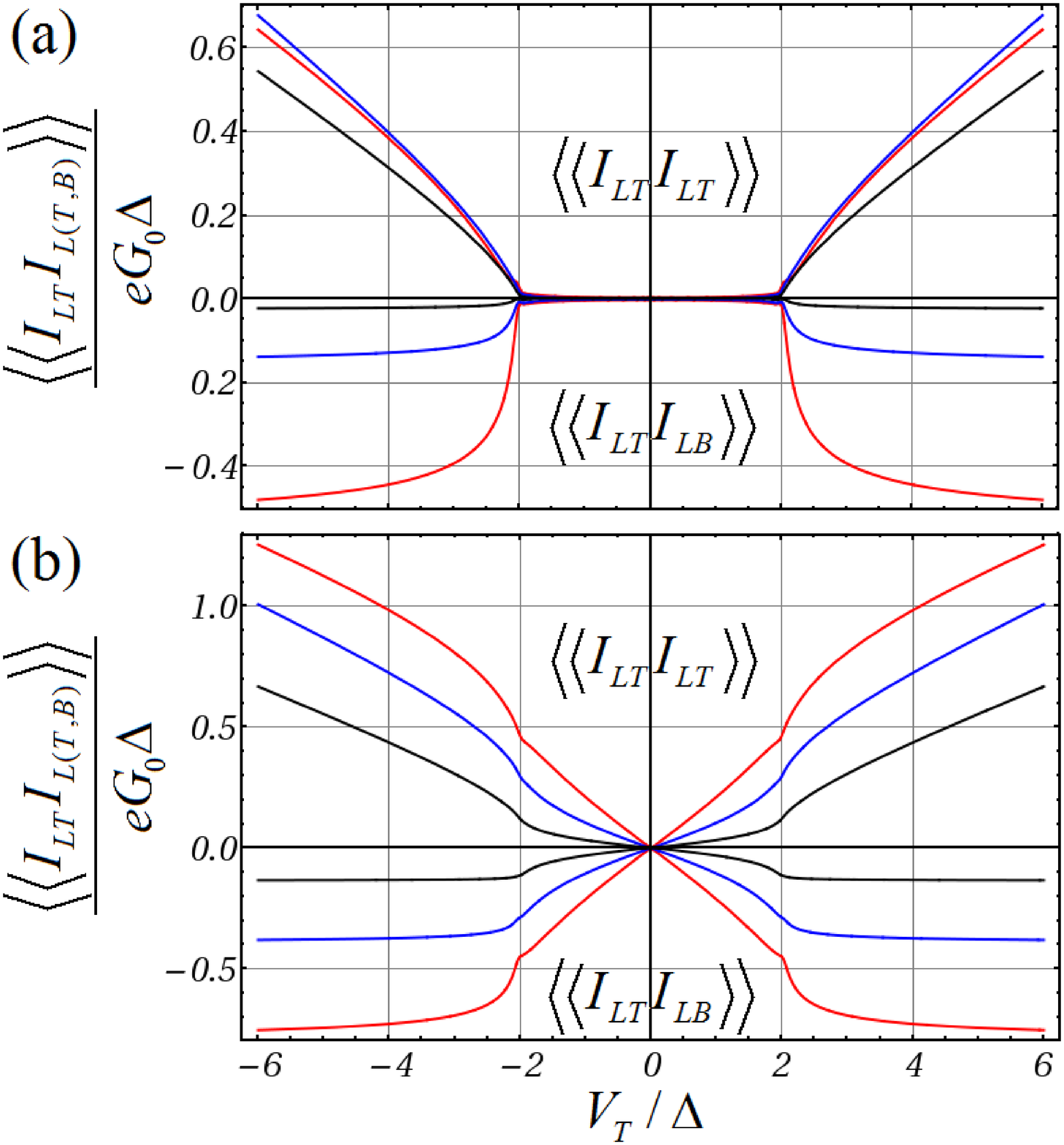}
\caption{(color online) (a) The parallel bias configuration (exciton blockade). Noise and cross-correlations are shown as a function of $V_T=V_B=V$, for the three values of transmission for the uncoupled system of $0.3$ (black), $0.5$ (blue), $0.7$ (red) and $T=0.01\Delta$. (b) The antiparallel bias configuration (counterpropagating currents). Noise and cross correlations are shown as a function of $V_T=-V_B=V$, again for the three values of contact transparency.}
\label{fig-noise}
\end{figure}

The current correlators are defined as $\cumu{I_{L \sigma}  I_{L \sigma^\prime}}   \doteq   \langle I_{\alpha \sigma} I_{\alpha^\prime \sigma^\prime} \rangle-\langle I_{\alpha \sigma} \rangle \langle I_{\alpha^\prime \sigma^\prime}\rangle $, and obtained from Eq.~(\ref{cgf}) as
%\begin{eqnarray*}
%\cumu{I_{\alpha \sigma}  I_{\alpha^\prime \sigma^\prime}}= (-i)^2 \frac{1}{\tau} \left. \frac{\partial^2}{\partial \lambda_{\alpha \sigma } \partial\lambda_{\alpha^\prime \sigma^\prime}} \ln \chi \right|_{\lambda_{\alpha \sigma} =0}  .
%\end{eqnarray*}
$\cumu{I_{\alpha \sigma}  I_{\alpha^\prime \sigma^\prime}}= (-i)^2  \tau^{-1} \left.  \partial^2 \ln \chi /\partial \lambda_{\alpha \sigma } \partial\lambda_{\alpha^\prime \sigma^\prime}  \right|_{\lambda_{\alpha \sigma} =0} $.
At equilibrium ($V_T=V_B=0$) we obtain the customary Johnson-Nyquist relation for $T\ll \Delta$
\begin{eqnarray}
\left. \cumu{I_{L T} I_{L T}}  \right|_{\mbox{eq}} &=& \left. -\cumu{ I_{L{T}} I_{R{T}}} \right|_{\mbox{eq}}=  4 T_A(0) G_0 k_B T \label{Seq}
\end{eqnarray}
This result indicates that the two electrons involved in an excitonic Andreev reflection dwell in separate layers so that only the conductance of a single layer enters the Johnson-Nyquist noise. The second equality in (\ref{Seq}) is
%reminiscent of
the generalized Johnson-Nyquist relationship obtained in [\onlinecite{PhysRevB.53.16390}] for a floating superconductor, which is here obtained without any use of Langevin forces. \\
%\fab{\bf From here on, I figured out an interpretation according to my personal understanding of the processes. However, to make sure that such interpretation is correct, comparison with plots is necessary. Please make some plots according to what is written in the text and in the caption of Fig.\ref{fig-noise}, and try to see if they match with the interpretation I gave. Would it be possible to make them for smaller temperature than $10^{-2} \Delta$?}
%The scenario is much richer in the out of equilibrium %situation. We shall focus in particular on
%, namely the case of equal and parallel  biases, $V_T=V_B$, (parallel configuration), and the case of the equal and opposite biases, $V_T=-V_B$, (antiparallel configuration).\\
As we see from Fig. \ref{fig-noise} also in nonequilibrium one always observes a negative cross-correlation. This is different from the case of a superconductor contacted to two normal electrodes, where one observes a positive cross-correlation of the two currents in the normal leads via crossed Andreev reflection.\cite{PhysRevB.63.165314,*2009arXiv0910.5558W,*19829377} The reason for this crucial difference is the fact that in the EC case we probe the correlation of electrons and holes rather then correlations of electron pairs as in the case of superconductors. We call the two voltage bias situations $V_T=\pm V_B$ parallel and antiparallel configurations. In the first configuration, where the average current is vanishing, the noise and the cross-correlation are shown in Fig.~\ref{fig-noise}(a). In the sub-gap regime, up to thermal fluctuation effects, both the noise and the cross-correlation vanish. This is because the incoming electrons and holes are always reflected back into the same lead they are injected from, since no exciton can penetrate inside the EC. Notice that this effect is essentially independent of the interface transmission, as shown in the three curves of Fig.~\ref{fig-noise}(a). Indeed for the parallel bias configuration in the subgap regime the EC gap effectively plays the role of a large barrier. In contrast, when $|V|>2 \Delta$, quasiparticles can be excited above the gap, and the noise in each lead starts deviating from zero, eventually increasing linearly with $V$.
%Notice, however, that the cross correlations are non-vanishing as well and saturate at a constant negative value. This can be understood by observing that each excitonic quasiparticle, both top-like and bottom-like, is a mixture of both top and bottom electron-hole excitations.  Thus, since current correlations at a voltage $V$ contain  the contributions from all energies up to $V$, both noise and cross correlations are increasing for range of voltages of a few $\Delta$s above the gap. For  $V\gg \Delta$, however, the weight of top (bottom) particle prevails for top-like (bottom-like) quasiparticles, implying that top and bottom layer become statistically independent. Thus, while the noise in each layer increases with voltage like in the non-interacting case, the cross correlation cannot further increase with $V$ [see Fig.~\ref{fig-noise}(a)] Notice also that for $V>2 \Delta$ the curves also depend on the interface transparency, since both electrons and holes can be transmitted via quasiparticles or be back reflected into the layer they are injected from. \\

In the antiparallel configuration, where the average currents in the layers flow in the opposite directions, the noise and the cross correlation are shown in Fig.~\ref{fig-noise}(b) as a function of $V=V_T=-V_B$. In the supra-gap regime the behavior is qualitatively similar to the parallel configuration, so that the noise increases and the cross-correlations saturates to a value determined by the quasiparticle mixed character above the gap. However, differences with respect to the parallel configuration emerge in the sub-gap regime, where both noise and cross correlation are now non-vanishing, and depend on the interface transmission.
\begin{figure}
\includegraphics[width=7cm]{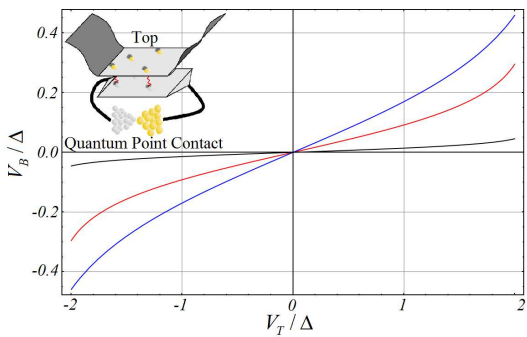}
\caption{(color online) Transformation of voltages using the EC. We show $V_{B}$ due to the current in the bottom layer coupled to a quantum point contact as a function of the applied voltage $V_{T}$ for different transparencies of the QPC,  $\Gamma_{L{T}} = 0.2 = \Gamma_{R{T}} = \Gamma_{L{B}} = \Gamma_{R{B}}$ and the QPC transparency is varied from $T_1 = 0.1$ (blue curve), $T_1 = 0.3$ (red) to $T_1 = 0.7$ (black). The inset shows the sketch of the experimental setup.}
\label{fig5}
\end{figure}

%\fab{\bf End of the part concerning the second cumulants}

%\vspace*{3cm}
%From FCS we also can obtain higher order current cumulants. In particular, the third cumulant follows the Levitov--Reznikov result \cite{PhysRevB.70.115305}
%\begin{eqnarray*}
%C_3 &=& \left. \frac{i}{\tau} \frac{\partial^3}{\partial \lambda_{L{T}}^3} \ln \chi_L(\lambda_{L{T}}, \lambda_{L{B}})\right|_{\lambda_{L{B}} = \lambda_{L{T}} = T =  0} \nonumber\\
%&=& e^2 \langle I_{L{T}} \rangle.
%\end{eqnarray*}

In realistic implementations of the proposed setup, a small single-particle inter-layer tunneling $t_{ab}$, here assumed vanishing, is expected to be present. The role of such a term has been discussed in the literature\cite{wen,*lozo1,*macdonald,*guseinov,sun2,pesin}, and depends on the specific implementation. In quantum Hall bilayers with GaAs wells, the role of such term is to lift the degeneracy of the EC phase and to introduce a finite critical current $I_c$,\cite{sun2} which depends on the layer area and on $|t_{ab}|^2$. In typical drag setups, the total flowing currents are usually higher than $I_c$.\cite{pesin} Inter-layer current is then negligible because the incoherent tunneling resistance is much higher than the in-plane resistance, so that the results presented here remain valid in the regime of currents $I_c \ll |I|$. We also checked that a long but finite exciton lifetime leading to a small imaginary part of $\Delta$ does not alter our results significantly. In the case of graphene bilayers, the scenario can be richer, depending on the nature of the insulating barrier and on the rotation of the graphene layers with respect to each other.

One possible application of the setup could be as a nanoscale voltage transformer. Ideal voltage transformers, such as inductors, require that the transformation coefficient  depends weakly on the load characteristics, and that energy losses are minimal. However, implementation of on-chip silicon based inductors turns out to be very difficult~\cite{cerofolini,*zolfaghari}, so that the seek for trade-off solutions in nanotransformers    represents a great challenge in modern electronics.
The  contact configuration  proposed in [\onlinecite{su2008}] suggests that
the setup can be used as a  voltage transformer at the nanoscale, contacting   the EC bottom layer   to another mesoscopic system with a known I-V characteristics, such as   a quantum point contact (QPC), where $I_{QPC} = G_0 T_1 V_{B}$, with $T_1$ being the contact transparency (see inset of Fig. \ref{fig5}).  At low temperatures  and for $|V_{{T}, \; {B}}| < 2\Delta$, only excitonic Andreev reflections contribute to transport and determine $\langle I_{L{T}} \rangle$ as a function of $V_{T}$.  On the other hand $\langle I_{L{T}} \rangle = - \langle I_{L{B}} \rangle$, and current conservation implies that $- \langle I_{L{B}} \rangle = I_{QPC}$. This leads to the transformation
\begin{eqnarray}
\langle I_{L{T}} \rangle (V_T)= G_0 T_1 V_{B}. \label{transform}
\end{eqnarray}
A typical example for the voltage interrelation is shown in Fig.~\ref{fig5}. In this case the transformation is controlled by $T_1$, which is typically tunable. Other realizations may involve transistors based on coupling to internal degrees of freedom. Importantly, nanotransformers based on the dissipationless EC counterflowing currents may help  minimizing heat and noise production.

%Although the density of states of an EC is highly nonlinear, there is also a rectifying impact of the linear I-V characteristics of the QPC.\\
To conclude, we have calculated the FCS for an EC contacted to four metallic leads. We have shown how counterpropagating currents and the generalized Johnson-Nyquist relation directly follow from the cumulant generating function. Using this approach we analyzed noise driven by excitonic Andreev reflections. Although the effective model is quadratic in fermion fields it correctly describes the non-trivial multi-particle exciton bound states. The lowest cumulants of charge transport resemble the free electron result, nonetheless fully accounting for the highly non-trivial drag effects as well as energy-dependence of the effective transmission coefficients. We also showed how excitons can be used for transforming current on the nanoscale.
The authors would like to thank S. Maier, R. Fazio, A.H. MacDonald, M. Polini, and F. Taddei  for many interesting discussions. The financial support was provided by the DFG under grant No. KO--2235/3, %by the Kompetenznetz "Funktionelle Nanostrukturen III" of the Baden-W\"{u}rttemberg Stiftung (Germany)
and `Enable fund' of the University of Heidelberg.


\begin{thebibliography}{43}%
\makeatletter
\providecommand \@ifxundefined [1]{%
 \@ifx{#1\undefined}
}%
\providecommand \@ifnum [1]{%
 \ifnum #1\expandafter \@firstoftwo
 \else \expandafter \@secondoftwo
 \fi
}%
\providecommand \@ifx [1]{%
 \ifx #1\expandafter \@firstoftwo
 \else \expandafter \@secondoftwo
 \fi
}%
\providecommand \natexlab [1]{#1}%
\providecommand \enquote  [1]{``#1''}%
\providecommand \bibnamefont  [1]{#1}%
\providecommand \bibfnamefont [1]{#1}%
\providecommand \citenamefont [1]{#1}%
\providecommand \href@noop [0]{\@secondoftwo}%
\providecommand \href [0]{\begingroup \@sanitize@url \@href}%
\providecommand \@href[1]{\@@startlink{#1}\@@href}%
\providecommand \@@href[1]{\endgroup#1\@@endlink}%
\providecommand \@sanitize@url [0]{\catcode `\\12\catcode `\$12\catcode
  `\&12\catcode `\#12\catcode `\^12\catcode `\_12\catcode `\%12\relax}%
\providecommand \@@startlink[1]{}%
\providecommand \@@endlink[0]{}%
\providecommand \url  [0]{\begingroup\@sanitize@url \@url }%
\providecommand \@url [1]{\endgroup\@href {#1}{\urlprefix }}%
\providecommand \urlprefix  [0]{URL }%
\providecommand \Eprint [0]{\href }%
\providecommand \doibase [0]{http://dx.doi.org/}%
\providecommand \selectlanguage [0]{\@gobble}%
\providecommand \bibinfo  [0]{\@secondoftwo}%
\providecommand \bibfield  [0]{\@secondoftwo}%
\providecommand \translation [1]{[#1]}%
\providecommand \BibitemOpen [0]{}%
\providecommand \bibitemStop [0]{}%
\providecommand \bibitemNoStop [0]{.\EOS\space}%
\providecommand \EOS [0]{\spacefactor3000\relax}%
\providecommand \BibitemShut  [1]{\csname bibitem#1\endcsname}%
\let\auto@bib@innerbib\@empty
%</preamble>
\bibitem [{\citenamefont {Keldysh}\ and\ \citenamefont
  {Kozlov}(1968)}]{keldysh1}%
  \BibitemOpen
  \bibfield  {author} {\bibinfo {author} {\bibfnamefont {L.~V.}\ \bibnamefont
  {Keldysh}}\ and\ \bibinfo {author} {\bibfnamefont {A.~N.}\ \bibnamefont
  {Kozlov}},\ }\href@noop {} {\bibfield  {journal} {\bibinfo  {journal} {Sov. Phys. JETP}\ }\textbf {\bibinfo
  {volume} {27}},\ \bibinfo {pages} {521} (\bibinfo {year} {1968})}\BibitemShut
  {NoStop}%
\bibitem [{\citenamefont {Eisenstein}\ and\ \citenamefont
  {MacDonald}(2004)}]{eisenstein1}%
  \BibitemOpen
  \bibfield  {author} {\bibinfo {author} {\bibfnamefont {J.~P.}\ \bibnamefont
  {Eisenstein}}\ and\ \bibinfo {author} {\bibfnamefont {A.~H.}\ \bibnamefont
  {MacDonald}},\ }\href@noop {} {\bibfield  {journal} {\bibinfo  {journal}
  {Nature}\ }\textbf {\bibinfo {volume} {432}},\ \bibinfo {pages} {691}
  (\bibinfo {year} {2004})}\BibitemShut {NoStop}%
\bibitem [{\citenamefont {Snoke}(2002)}]{snoke}%
  \BibitemOpen
  \bibfield  {author} {\bibinfo {author} {\bibfnamefont {D.}~\bibnamefont
  {Snoke}},\ }\href@noop {} {\bibfield  {journal} {\bibinfo  {journal}
  {Science}\ }\textbf {\bibinfo {volume} {298}},\ \bibinfo {pages} {1368}
  (\bibinfo {year} {2002})}\BibitemShut {NoStop}%
\bibitem [{\citenamefont {Eisenstein}\ \emph {et~al.}(1990)\citenamefont
  {Eisenstein}, \citenamefont {Pfeiffer},\ and\ \citenamefont
  {West}}]{eisenstein2}%
  \BibitemOpen
  \bibfield  {author} {\bibinfo {author} {\bibfnamefont {J.~P.}\ \bibnamefont
  {Eisenstein}}, \bibinfo {author} {\bibfnamefont {L.~N.}\ \bibnamefont
  {Pfeiffer}}, \ and\ \bibinfo {author} {\bibfnamefont {K.~W.}\ \bibnamefont
  {West}},\ }\href@noop {} {\bibfield  {journal} {\bibinfo  {journal} {Appl.
  Phys. Lett.}\ }\textbf {\bibinfo {volume} {57}},\ \bibinfo {pages} {2324}
  (\bibinfo {year} {1990})}\BibitemShut {NoStop}%
\bibitem [{\citenamefont {Zhou}\ and\ \citenamefont
  {Kim}(1999)}]{PhysRevB.59.R7825}%
  \BibitemOpen
  \bibfield  {author} {\bibinfo {author} {\bibfnamefont {F.}~\bibnamefont
  {Zhou}}\ and\ \bibinfo {author} {\bibfnamefont {Y.~B.}\ \bibnamefont {Kim}},\
  }\href {\doibase 10.1103/PhysRevB.59.R7825} {\bibfield  {journal} {\bibinfo
  {journal} {Phys. Rev. B}\ }\textbf {\bibinfo {volume} {59}},\ \bibinfo
  {pages} {R7825} (\bibinfo {year} {1999})}\BibitemShut {NoStop}%
\bibitem [{\citenamefont {Hu}(2000)}]{PhysRevLett.85.820}%
  \BibitemOpen
  \bibfield  {author} {\bibinfo {author} {\bibfnamefont{Ben-YuKuang} \bibnamefont{Hu}},\
  }\href {\doibase 10.1103/PhysRevLett.85.820} {\bibfield
  {journal} {\bibinfo  {journal} {Phys. Rev. Lett.}\ }\textbf {\bibinfo
  {volume} {85}},\ \bibinfo {pages} {820} (\bibinfo {year} {2000})}\BibitemShut
  {NoStop}%
\bibitem [{\citenamefont {Eisenstein}\ \emph {et~al.}(1992)\citenamefont
  {Eisenstein}, \citenamefont {Boebinger}, \citenamefont {Pfeiffer},
  \citenamefont {West},\ and\ \citenamefont {He}}]{PhysRevLett.68.1383}%
  \BibitemOpen
  \bibfield  {author} {\bibinfo {author} {\bibfnamefont {J.~P.}\ \bibnamefont
  {Eisenstein}}, \bibinfo {author} {\bibfnamefont {G.~S.}\ \bibnamefont
  {Boebinger}}, \bibinfo {author} {\bibfnamefont {L.~N.}\ \bibnamefont
  {Pfeiffer}}, \emph {et~al.},\ }\href
  {\doibase 10.1103/PhysRevLett.68.1383} {\bibfield  {journal} {\bibinfo
  {journal} {Phys. Rev. Lett.}\ }\textbf {\bibinfo {volume} {68}},\ \bibinfo
  {pages} {1383} (\bibinfo {year} {1992})}\BibitemShut {NoStop}%
\bibitem [{\citenamefont {Lilly}\ \emph {et~al.}(1998)\citenamefont {Lilly},
  \citenamefont {Eisenstein}, \citenamefont {Pfeiffer},\ and\ \citenamefont
  {West}}]{PhysRevLett.80.1714}%
  \BibitemOpen
  \bibfield  {author} {\bibinfo {author} {\bibfnamefont {M.~P.}\ \bibnamefont
  {Lilly}}, \bibinfo {author} {\bibfnamefont {J.~P.}\ \bibnamefont
  {Eisenstein}}, \bibinfo {author} {\bibfnamefont {L.~N.}\ \bibnamefont
  {Pfeiffer}}, \emph {et~al.},\ }\href {\doibase 10.1103/PhysRevLett.80.1714} {\bibfield  {journal}
  {\bibinfo  {journal} {Phys. Rev. Lett.}\ }\textbf {\bibinfo {volume} {80}},\
  \bibinfo {pages} {1714} (\bibinfo {year} {1998})}\BibitemShut {NoStop}%
\bibitem [{\citenamefont {Spielman}\ \emph {et~al.}(2000)\citenamefont
  {Spielman}, \citenamefont {Eisenstein}, \citenamefont {Pfeiffer},\ and\
  \citenamefont {West}}]{PhysRevLett.84.5808}%
  \BibitemOpen
  \bibfield  {author} {\bibinfo {author} {\bibfnamefont {I.~B.}\ \bibnamefont
  {Spielman}}, \bibinfo {author} {\bibfnamefont {J.~P.}\ \bibnamefont
  {Eisenstein}}, \bibinfo {author} {\bibfnamefont {L.~N.}\ \bibnamefont
  {Pfeiffer}}, \emph {et~al.},\ }\href {\doibase 10.1103/PhysRevLett.84.5808} {\bibfield  {journal}
  {\bibinfo  {journal} {Phys. Rev. Lett.}\ }\textbf {\bibinfo {volume} {84}},\
  \bibinfo {pages} {5808} (\bibinfo {year} {2000})}\BibitemShut {NoStop}%
\bibitem [{\citenamefont {Kellogg}\ \emph {et~al.}(2004)\citenamefont
  {Kellogg}, \citenamefont {Eisenstein}, \citenamefont {Pfeiffer},\ and\
  \citenamefont {West}}]{PhysRevLett.93.036801}%
  \BibitemOpen
  \bibfield  {author} {\bibinfo {author} {\bibfnamefont {M.}~\bibnamefont
  {Kellogg}}, \bibinfo {author} {\bibfnamefont {J.~P.}\ \bibnamefont
  {Eisenstein}}, \bibinfo {author} {\bibfnamefont {L.~N.}\ \bibnamefont
  {Pfeiffer}}, \emph {et~al.},\ }\href {\doibase 10.1103/PhysRevLett.93.036801} {\bibfield
  {journal} {\bibinfo  {journal} {Phys. Rev. Lett.}\ }\textbf {\bibinfo
  {volume} {93}},\ \bibinfo {pages} {036801} (\bibinfo {year}
  {2004})}\BibitemShut {NoStop}%
\bibitem [{\citenamefont {Tutuc}\ \emph {et~al.}(2004)\citenamefont {Tutuc},
  \citenamefont {Shayegan},\ and\ \citenamefont
  {Huse}}]{PhysRevLett.93.036802}%
  \BibitemOpen
  \bibfield  {author} {\bibinfo {author} {\bibfnamefont {E.}~\bibnamefont
  {Tutuc}}, \bibinfo {author} {\bibfnamefont {M.}~\bibnamefont {Shayegan}}, \
  and\ \bibinfo {author} {\bibfnamefont {D.~A.}\ \bibnamefont {Huse}},\ }\href
  {\doibase 10.1103/PhysRevLett.93.036802} {\bibfield  {journal} {\bibinfo
  {journal} {Phys. Rev. Lett.}\ }\textbf {\bibinfo {volume} {93}},\ \bibinfo
  {pages} {036802} (\bibinfo {year} {2004})}\BibitemShut {NoStop}%
\bibitem [{\citenamefont {Tiemann}\ \emph {et~al.}(2008)\citenamefont
  {Tiemann}, \citenamefont {Dietsche}, \citenamefont {Hauser},\ and\
  \citenamefont {v.~Klitzing}}]{1367-2630-10-4-045018}%
  \BibitemOpen
  \bibfield  {author} {\bibinfo {author} {\bibfnamefont {L.}~\bibnamefont
  {Tiemann}}, \bibinfo {author} {\bibfnamefont {W.}~\bibnamefont {Dietsche}},
  \bibinfo {author} {\bibfnamefont {M.}~\bibnamefont {Hauser}}, \emph {et~al.},\ }\href@noop {}
  {\bibfield  {journal} {\bibinfo  {journal} {New J. Phys.}\ }\textbf {\bibinfo
  {volume} {10}},\ \bibinfo {pages} {045018} (\bibinfo {year}
  {2008})}\BibitemShut {NoStop}%
\bibitem [{\citenamefont {Finck}\ \emph {et~al.}(2011)\citenamefont {Finck},
  \citenamefont {Eisenstein}, \citenamefont {Pfeiffer},\ and\ \citenamefont
  {West}}]{PhysRevLett.106.236807}%
  \BibitemOpen
  \bibfield  {author} {\bibinfo {author} {\bibfnamefont {A.~D.~K.}\
  \bibnamefont {Finck}}, \bibinfo {author} {\bibfnamefont {J.~P.}\ \bibnamefont
  {Eisenstein}}, \bibinfo {author} {\bibfnamefont {L.~N.}\ \bibnamefont
  {Pfeiffer}}, \emph {et~al.},\ }\href {\doibase 10.1103/PhysRevLett.106.236807} {\bibfield
  {journal} {\bibinfo  {journal} {Phys. Rev. Lett.}\ }\textbf {\bibinfo
  {volume} {106}},\ \bibinfo {pages} {236807} (\bibinfo {year}
  {2011})}\BibitemShut {NoStop}%
\bibitem [{\citenamefont {Croxall}\ \emph {et~al.}(2008)\citenamefont
  {Croxall}, \citenamefont {Das~Gupta}, \citenamefont {Nicoll}, \citenamefont
  {Thangaraj}, \citenamefont {Beere}, \citenamefont {Farrer}, \citenamefont
  {Ritchie},\ and\ \citenamefont {Pepper}}]{PhysRevLett.101.246801}%
  \BibitemOpen
  \bibfield  {author} {\bibinfo {author} {\bibfnamefont {A.~F.}\ \bibnamefont
  {Croxall}}, \bibinfo {author} {\bibfnamefont {K.}~\bibnamefont {Das~Gupta}},
  \bibinfo {author} {\bibfnamefont {C.~A.}\ \bibnamefont {Nicoll}}, \emph {et~al.},\ }\href {\doibase
  10.1103/PhysRevLett.101.246801} {\bibfield  {journal} {\bibinfo  {journal}
  {Phys. Rev. Lett.}\ }\textbf {\bibinfo {volume} {101}},\ \bibinfo {pages}
  {246801} (\bibinfo {year} {2008})}\BibitemShut {NoStop}%
\bibitem [{\citenamefont {Seamons}\ \emph {et~al.}(2009)\citenamefont
  {Seamons}, \citenamefont {Morath}, \citenamefont {Reno},\ and\ \citenamefont
  {Lilly}}]{PhysRevLett.102.026804}%
  \BibitemOpen
  \bibfield  {author} {\bibinfo {author} {\bibfnamefont {J.~A.}\ \bibnamefont
  {Seamons}}, \bibinfo {author} {\bibfnamefont {C.~P.}\ \bibnamefont {Morath}},
  \bibinfo {author} {\bibfnamefont {J.~L.}\ \bibnamefont {Reno}}, \emph {et~al.},\ }\href
  {\doibase 10.1103/PhysRevLett.102.026804} {\bibfield  {journal} {\bibinfo
  {journal} {Phys. Rev. Lett.}\ }\textbf {\bibinfo {volume} {102}},\ \bibinfo
  {pages} {026804} (\bibinfo {year} {2009})}\BibitemShut {NoStop}%
\bibitem [{\citenamefont {Su}\ and\ \citenamefont {MacDonald}(2008)}]{su2008}%
  \BibitemOpen
  \bibfield  {author} {\bibinfo {author} {\bibfnamefont {J.}~\bibnamefont
  {Su}}\ and\ \bibinfo {author} {\bibfnamefont {A.~H.}\ \bibnamefont
  {MacDonald}},\ }\href@noop {} {\bibfield  {journal} {\bibinfo  {journal}
  {Nat. Phys.}\ }\textbf {\bibinfo {volume} {4}},\ \bibinfo {pages} {799}
  (\bibinfo {year} {2008})}\BibitemShut {NoStop}%
\bibitem [{\citenamefont {{Mink}}\ \emph {et~al.}(2011)\citenamefont {{Mink}},
  \citenamefont {{Stoof}}, \citenamefont {{Duine}}, \citenamefont {{Polini}},\
  and\ \citenamefont {{Vignale}}}]{2011arXiv1108.2298M}%
  \BibitemOpen
  \bibfield  {author} {\bibinfo {author} {\bibfnamefont {M.~P.}\ \bibnamefont
  {{Mink}}}, \bibinfo {author} {\bibfnamefont {H.~T.~C.}\ \bibnamefont
  {{Stoof}}}, \bibinfo {author} {\bibfnamefont {R.~A.}\ \bibnamefont
  {{Duine}}}, \emph {et~al.},\
  }\href@noop {} {\bibfield  {journal} {\bibinfo  {journal} {ArXiv e-prints}\ }
  (\bibinfo {year} {2011})},\ \Eprint {http://arxiv.org/abs/1108.2298}
  {arXiv:1108.2298 [cond-mat.mes-hall]} \BibitemShut {NoStop}%
\bibitem [{\citenamefont {Dolcini}\ \emph {et~al.}(2010)\citenamefont
  {Dolcini}, \citenamefont {Rainis}, \citenamefont {Taddei}, \citenamefont
  {Polini}, \citenamefont {Fazio},\ and\ \citenamefont
  {MacDonald}}]{PhysRevLett.104.027004}%
  \BibitemOpen
  \bibfield  {author} {\bibinfo {author} {\bibfnamefont {F.}~\bibnamefont
  {Dolcini}}, \bibinfo {author} {\bibfnamefont {D.}~\bibnamefont {Rainis}},
  \bibinfo {author} {\bibfnamefont {F.}~\bibnamefont {Taddei}}, \emph {et~al.},\  }\href {\doibase
  10.1103/PhysRevLett.104.027004} {\bibfield  {journal} {\bibinfo  {journal}
  {Phys. Rev. Lett.}\ }\textbf {\bibinfo {volume} {104}},\ \bibinfo {pages}
  {027004} (\bibinfo {year} {2010})}\BibitemShut {NoStop}%
\bibitem [{\citenamefont {Peotta}\ \emph {et~al.}(2011)\citenamefont
  {Peotta}, \citenamefont {{Gibertini}}, \citenamefont {{Dolcini}},
  \citenamefont {{Taddei}}, \citenamefont {{Polini}}, \citenamefont {{Ioffe}},
  \citenamefont {{Fazio}},\ and\ \citenamefont
  {{MacDonald}}}]{2011arXiv1108.1533P}%
  \BibitemOpen
  \bibfield  {author} {\bibinfo {author} {\bibfnamefont {S.}~\bibnamefont
  {Peotta}}, \bibinfo {author} {\bibfnamefont {M.}~\bibnamefont
  {Gilbertini}}, \bibinfo {author} {\bibfnamefont {F.}~\bibnamefont
  {Dolcini}}, \emph {et~al.},\ }\href {\doibase
  10.1103/PhysRevB.84.184528} {\bibfield  {journal} {\bibinfo  {journal}
  {Phys. Rev. B}\ }\textbf {\bibinfo {volume} {84}},\ \bibinfo {pages}
  {184528} (\bibinfo {year} {2011})}\BibitemShut {NoStop}%
\bibitem [{\citenamefont {Narozhny}\ and\ \citenamefont
  {Aleiner}(2000)}]{PhysRevLett.84.5383}%
  \BibitemOpen
  \bibfield  {author} {\bibinfo {author} {\bibfnamefont {B.~N.}\ \bibnamefont
  {Narozhny}}\ and\ \bibinfo {author} {\bibfnamefont {I.~L.}\ \bibnamefont
  {Aleiner}},\ }\href {\doibase 10.1103/PhysRevLett.84.5383} {\bibfield
  {journal} {\bibinfo  {journal} {Phys. Rev. Lett.}\ }\textbf {\bibinfo
  {volume} {84}},\ \bibinfo {pages} {5383} (\bibinfo {year}
  {2000})}\BibitemShut {NoStop}%
\bibitem [{\citenamefont {Mortensen}\ \emph {et~al.}(2002)\citenamefont
  {Mortensen}, \citenamefont {Flensberg},\ and\ \citenamefont
  {Jauho}}]{PhysRevB.65.085317}%
  \BibitemOpen
  \bibfield  {author} {\bibinfo {author} {\bibfnamefont {N.~A.}\ \bibnamefont
  {Mortensen}}, \bibinfo {author} {\bibfnamefont {K.}~\bibnamefont
  {Flensberg}}, \ and\ \bibinfo {author} {\bibfnamefont {A.-P.}\ \bibnamefont
  {Jauho}},\ }\href {\doibase 10.1103/PhysRevB.65.085317} {\bibfield  {journal}
  {\bibinfo  {journal} {Phys. Rev. B}\ }\textbf {\bibinfo {volume} {65}},\
  \bibinfo {pages} {085317} (\bibinfo {year} {2002})}\BibitemShut {NoStop}%
\bibitem [{\citenamefont {Min}\ \emph {et~al.}(2008)\citenamefont {Min},
  \citenamefont {Bistritzer}, \citenamefont {Su}, \citenamefont {MacDonald}}]{PhysRevB.78.121401}%
  \BibitemOpen
  \bibfield  {author} {\bibinfo {author} {\bibfnamefont {H.}~\bibnamefont
  {Min}}, \bibinfo {author} {\bibfnamefont {R.}~\bibnamefont {Bistritzer}},
  \bibinfo {author} {\bibfnamefont {J.-J.}\ \bibnamefont {Su}}, \ and\ \bibinfo
  {author} {\bibfnamefont {A.~H.}\ \bibnamefont {MacDonald}},\ }\href {\doibase
  10.1103/PhysRevB.78.121401} {\bibfield  {journal} {\bibinfo  {journal} {Phys.
  Rev. B}\ }\textbf {\bibinfo {volume} {78}},\ \bibinfo {pages} {121401}
  (\bibinfo {year} {2008})}\BibitemShut {NoStop}%
\bibitem [{\citenamefont {Lozovik}\ and\ \citenamefont
  {Sokolik}(2008)}]{lozovik2008}%
  \BibitemOpen
  \bibfield  {author} {\bibinfo {author} {\bibfnamefont {Yu.~E.}\ \bibnamefont
  {Lozovik}}\ and\ \bibinfo {author} {\bibfnamefont {A.~A.}\ \bibnamefont
  {Sokolik}},\ }\href@noop {} {\bibfield  {journal} {\bibinfo  {journal} {JETP
  Lett.}\ }\textbf {\bibinfo {volume} {87}},\ \bibinfo {pages} {55} (\bibinfo
  {year} {2008})}\BibitemShut {NoStop}%
\bibitem [{\citenamefont {{Ponomarenko}}\ \emph {et~al.}(2011)\citenamefont
  {{Ponomarenko}}, \citenamefont {{Zhukov}}, \citenamefont {{Jalil}},
  \citenamefont {{Morozov}}, \citenamefont {{Novoselov}}, \citenamefont
  {{Cheianov}}, \citenamefont {{Fal'ko}}, \citenamefont {{Watanabe}},
  \citenamefont {{Taniguchi}}, \citenamefont {{Geim}},\ and\ \citenamefont
  {{Gorbachev}}}]{2011arXiv1107.0115P}%
  \BibitemOpen
  \bibfield  {author} {\bibinfo {author} {\bibfnamefont {L.~A.}\ \bibnamefont
  {{Ponomarenko}}}, \bibinfo {author} {\bibfnamefont {A.~A.}\ \bibnamefont
  {{Zhukov}}}, \bibinfo {author} {\bibfnamefont {R.}~\bibnamefont {{Jalil}}}, \emph {et~al.},\ }\href@noop {}
  {\bibfield  {journal} {\bibinfo  {journal} {ArXiv e-prints}\ } (\bibinfo
  {year} {2011})},\ \Eprint {http://arxiv.org/abs/1107.0115} {arXiv:1107.0115
  [cond-mat.mes-hall]} \BibitemShut {NoStop}%
\bibitem [{\citenamefont {Ivanov}\ and\ \citenamefont
  {Levitov}(1993)}]{levitov1}%
  \BibitemOpen
  \bibfield  {author} {\bibinfo {author} {\bibfnamefont {D.~A.}\ \bibnamefont
  {Ivanov}}\ and\ \bibinfo {author} {\bibfnamefont {L.~S.}\ \bibnamefont
  {Levitov}},\ }\href@noop {} {\bibfield  {journal} {\bibinfo  {journal} {JETP
  Lett.}\ }\textbf {\bibinfo {volume} {58}},\ \bibinfo {pages} {461} (\bibinfo
  {year} {1993})}\BibitemShut {NoStop}%
\bibitem [{\citenamefont {Nazarov}(1999)}]{nazarov}%
  \BibitemOpen
  \bibfield  {author} {\bibinfo {author} {\bibfnamefont {Yu.~V.}\ \bibnamefont
  {Nazarov}},\ }\href@noop {} {\bibfield  {journal} {\bibinfo  {journal} {Ann.
  Phys. (Leipzig)}\ }\textbf {\bibinfo {volume} {8}},\ \bibinfo {pages} {507
  – 510} (\bibinfo {year} {1999})}\BibitemShut {NoStop}%
\bibitem [{\citenamefont {Nazarov}\ and\ \citenamefont
  {Kindermann}(2003)}]{kindermann}%
  \BibitemOpen
  \bibfield  {author} {\bibinfo {author} {\bibfnamefont {Yu.~V.}\ \bibnamefont
  {Nazarov}}\ and\ \bibinfo {author} {\bibfnamefont {M.}~\bibnamefont
  {Kindermann}},\ }\href@noop {} {\bibfield  {journal} {\bibinfo  {journal}
  {Eur. Phys. J. B}\ }\textbf {\bibinfo {volume} {35}},\ \bibinfo {pages} {413}
  (\bibinfo {year} {2003})}\BibitemShut {NoStop}%
\bibitem [{\citenamefont {Sch\"on}(1981)}]{Gerd1981171}%
  \BibitemOpen
  \bibfield  {author} {\bibinfo {author} {\bibfnamefont {G.}~\bibnamefont
  {Sch\"on}},\ }\href {\doibase 10.1016/0378-4363(81)90392-2} {\bibfield
  {journal} {\bibinfo  {journal} {Physica B+C}\ }\textbf {\bibinfo {volume}
  {107}},\ \bibinfo {pages} {171 } (\bibinfo {year} {1981})}\BibitemShut
  {NoStop}%
\bibitem [{\citenamefont {Schmidt}\ \emph {et~al.}(2007)\citenamefont
  {Schmidt}, \citenamefont {Komnik},\ and\ \citenamefont
  {Gogolin}}]{PhysRevLett.98.056603}%
  \BibitemOpen
  \bibfield  {author} {\bibinfo {author} {\bibfnamefont {T.~L.}\ \bibnamefont
  {Schmidt}}, \bibinfo {author} {\bibfnamefont {A.}~\bibnamefont {Komnik}}, \
  and\ \bibinfo {author} {\bibfnamefont {A.~O.}\ \bibnamefont {Gogolin}},\
  }\href {\doibase 10.1103/PhysRevLett.98.056603} {\bibfield  {journal}
  {\bibinfo  {journal} {Phys. Rev. Lett.}\ }\textbf {\bibinfo {volume} {98}},\
  \bibinfo {pages} {056603} (\bibinfo {year} {2007})}\BibitemShut {NoStop}%
\bibitem [{\citenamefont {Gogolin}\ and\ \citenamefont
  {Komnik}(2006)}]{PhysRevB.73.195301}%
  \BibitemOpen
  \bibfield  {author} {\bibinfo {author} {\bibfnamefont {A.~O.}\ \bibnamefont
  {Gogolin}}\ and\ \bibinfo {author} {\bibfnamefont {A.}~\bibnamefont
  {Komnik}},\ }\href {\doibase 10.1103/PhysRevB.73.195301} {\bibfield
  {journal} {\bibinfo  {journal} {Phys. Rev. B}\ }\textbf {\bibinfo {volume}
  {73}},\ \bibinfo {pages} {195301} (\bibinfo {year} {2006})}\BibitemShut
  {NoStop}%
\bibitem [{\citenamefont {Cuevas}\ \emph {et~al.}(1996)\citenamefont {Cuevas},
  \citenamefont {Mart\'\i{}n-Rodero},\ and\ \citenamefont
  {Levy Yeyati}}]{PhysRevB.54.7366}%
  \BibitemOpen
  \bibfield  {author} {\bibinfo {author} {\bibfnamefont {J.~C.}\ \bibnamefont
  {Cuevas}}, \bibinfo {author} {\bibfnamefont {A.}~\bibnamefont
  {Mart\'\i{}n-Rodero}}, \ and\ \bibinfo {author} {\bibfnamefont {A.}\
  \bibnamefont {Levy Yeyati}},\ }\href {\doibase 10.1103/PhysRevB.54.7366}
  {\bibfield  {journal} {\bibinfo  {journal} {Phys. Rev. B}\ }\textbf {\bibinfo
  {volume} {54}},\ \bibinfo {pages} {7366} (\bibinfo {year}
  {1996})}\BibitemShut {NoStop}%
\bibitem [{Note1()}]{Note1}%
  \BibitemOpen
  \bibinfo {note} {In principle a finite $q$ also affects the gap. We neglect
  this effect here, as customary in similar situations concerning
  superconductors.}\BibitemShut {Stop}%
\bibitem [{\citenamefont {Lambert}(1991)}]{lambert}%
  \BibitemOpen
  \bibfield  {author} {\bibinfo {author} {\bibfnamefont {C.~J.}\ \bibnamefont
  {Lambert}},\ }\href@noop {} {\bibfield  {journal} {\bibinfo  {journal} {J.
  Phys.: Condens. Matter}\ }\textbf {\bibinfo {volume} {3}},\ \bibinfo {pages}
  {6579} (\bibinfo {year} {1991})}\BibitemShut {NoStop}%
\bibitem [{\citenamefont {Martin}\ and\ \citenamefont
  {Lambert}(1995)}]{PhysRevB.51.17999}%
  \BibitemOpen
  \bibfield  {author} {\bibinfo {author} {\bibfnamefont {A.}~\bibnamefont
  {Martin}}\ and\ \bibinfo {author} {\bibfnamefont {C.~J.}\ \bibnamefont
  {Lambert}},\ }\href {\doibase 10.1103/PhysRevB.51.17999} {\bibfield
  {journal} {\bibinfo  {journal} {Phys. Rev. B}\ }\textbf {\bibinfo {volume}
  {51}},\ \bibinfo {pages} {17999} (\bibinfo {year} {1995})}\BibitemShut
  {NoStop}%
\bibitem [{\citenamefont {Anantram}\ and\ \citenamefont
  {Datta}(1996)}]{PhysRevB.53.16390}%
  \BibitemOpen
  \bibfield  {author} {\bibinfo {author} {\bibfnamefont {M.~P.}\ \bibnamefont
  {Anantram}}\ and\ \bibinfo {author} {\bibfnamefont {S.}~\bibnamefont
  {Datta}},\ }\href {\doibase 10.1103/PhysRevB.53.16390} {\bibfield  {journal}
  {\bibinfo  {journal} {Phys. Rev. B}\ }\textbf {\bibinfo {volume} {53}},\
  \bibinfo {pages} {16390} (\bibinfo {year} {1996})}\BibitemShut {NoStop}%
\bibitem [{\citenamefont {Recher}\ \emph {et~al.}(2001)\citenamefont {Recher},
  \citenamefont {Sukhorukov},\ and\ \citenamefont {Loss}}]{PhysRevB.63.165314}%
  \BibitemOpen
  \bibfield  {author} {\bibinfo {author} {\bibfnamefont {P.}~\bibnamefont
  {Recher}}, \bibinfo {author} {\bibfnamefont {E.~V.}\ \bibnamefont
  {Sukhorukov}}, \ and\ \bibinfo {author} {\bibfnamefont {D.}~\bibnamefont
  {Loss}},\ }\href {\doibase 10.1103/PhysRevB.63.165314} {\bibfield  {journal}
  {\bibinfo  {journal} {Phys. Rev. B}\ }\textbf {\bibinfo {volume} {63}},\
  \bibinfo {pages} {165314} (\bibinfo {year} {2001})}\BibitemShut {NoStop}%
\bibitem [{\citenamefont {{Wei}}\ and\ \citenamefont
  {{Chandrasekhar}}(2010)}]{2009arXiv0910.5558W}%
  \BibitemOpen
  \bibfield  {author} {\bibinfo {author} {\bibfnamefont {J.}~\bibnamefont
  {{Wei}}}\ and\ \bibinfo {author} {\bibfnamefont {V.}~\bibnamefont
  {{Chandrasekhar}}},\ }\href@noop {} {\bibfield  {journal} {\bibinfo
  {journal} {Nat. Phys.}\ }\textbf {\bibinfo {volume} {6}},\ \bibinfo {pages}
  {494} (\bibinfo {year} {2010})}\BibitemShut {NoStop}%
\bibitem [{\citenamefont {Hofstetter}\ \emph {et~al.}(2009)\citenamefont
  {Hofstetter}, \citenamefont {Csonka}, \citenamefont {Nyg\r{a}rd},\ and\
  \citenamefont {Sch\"{o}nenberger}}]{19829377}%
  \BibitemOpen
  \bibfield  {author} {\bibinfo {author} {\bibfnamefont {L.}~\bibnamefont
  {Hofstetter}}, \bibinfo {author} {\bibfnamefont {S.}~\bibnamefont {Csonka}},
  \bibinfo {author} {\bibfnamefont {J.}~\bibnamefont {Nyg\r{a}rd}}, \emph {et~al.},\
  }\href@noop {} {\bibfield  {journal} {\bibinfo  {journal} {Nature}\ }\textbf
  {\bibinfo {volume} {461}},\ \bibinfo {pages} {960} (\bibinfo {year}
  {2009})}\BibitemShut {NoStop}%
\bibitem [{\citenamefont {{Cerofolini}}(2011)}]{cerofolini}%
  \BibitemOpen
  \bibfield  {author} {\bibinfo {author} {\bibfnamefont {G.~F.}~\bibnamefont
  {{Cerofolini}}},\ }\href@noop {} {\bibfield  {journal} {\bibinfo
  {journal} {Appl. Phys. A}\ } \bibinfo {pages}
  {1-16} (\bibinfo {year} {2011})}\BibitemShut {NoStop}%
\bibitem [{\citenamefont {Zolfaghari}\ \emph {et~al.}(2001)\citenamefont
  {Zolfaghari}, \citenamefont {Chan}, \ and\
  \citenamefont {Razavi}}]{zolfaghari}%
  \BibitemOpen
  \bibfield  {author} {\bibinfo {author} {\bibfnamefont {A.}~\bibnamefont
  {Zolfaghari}}, \bibinfo {author} {\bibfnamefont {A.}~\bibnamefont {Chan}},
  \ and\
  \bibinfo {author} {\bibfnamefont {B.}~\bibnamefont {Razavi}},\
  }\href@noop {} {\bibfield  {journal} {\bibinfo  {journal} {IEEE J. Solid-State Circuits}\ }\textbf
  {\bibinfo {volume} {36}},\ \bibinfo {pages} {620} (\bibinfo {year}
  {2001})}\BibitemShut {NoStop}%
\bibitem [{\citenamefont {Wen}(1993)}]{wen}%
  \BibitemOpen
  \bibfield  {author} {\bibinfo {author} {\bibfnamefont {X.~G.}\ \bibnamefont
  {Wen}}, \ and\ \bibinfo {author} {\bibfnamefont {A.}\
  \bibnamefont {Zee}}, \ }\href@noop {} {\bibfield  {journal} {\bibinfo  {journal} {Phys. Rev. B}\ }\textbf {\bibinfo {volume} {47}},\ \bibinfo {pages} {2265} (\bibinfo {year} {1993})}\BibitemShut {NoStop}%
\bibitem [{\citenamefont {Lozovik}(1977)}]{lozo1}%
  \BibitemOpen
  \bibfield  {author} {\bibinfo {author} {\bibfnamefont {Yu.~E.}\ \bibnamefont
  {Lozovik}}, \ and\ \bibinfo {author} {\bibfnamefont {V.~I.}\ \bibnamefont
  {Yudson}} , \ }\href@noop {} {\bibfield  {journal} {\bibinfo  {journal} {JETP Lett.}\ }\textbf {\bibinfo {volume} {25}},\ \bibinfo {pages} {14} (\bibinfo {year} {1977})}\BibitemShut {NoStop}%
\bibitem [{\citenamefont {MacDonald}\ and\ \citenamefont
  {Platzmann}(1990)}]{macdonald}%
  \BibitemOpen
  \bibfield  {author} {\bibinfo {author} {\bibfnamefont {A.~H.}\ \bibnamefont
  {MacDonald}}, \bibinfo {author} {\bibfnamefont {P.~M.}\ \bibnamefont
  {Platzman}}, \ and\ \bibinfo {author} {\bibfnamefont {G.~S.}~\bibnamefont
  {Boebinger}},\ }\href@noop {} {\bibfield  {journal} {\bibinfo  {journal}
  {Phys. Rev. Lett.}\ }\textbf {\bibinfo {volume} {65}},\ \bibinfo {pages} {775}
  (\bibinfo {year} {1990})}\BibitemShut {NoStop}%
\bibitem [{\citenamefont {Guseinov}\ and\ \citenamefont
  {Keldysh}(1973)}]{guseinov}%
  \BibitemOpen
  \bibfield  {author} {\bibinfo {author} {\bibfnamefont {R.~R.}\ \bibnamefont
  {Guseinov}}, \ and\ \bibinfo {author} {\bibfnamefont {L.~V.}~\bibnamefont
  {Keldysh}},\ }\href@noop {} {\bibfield  {journal} {\bibinfo  {journal}
  {Sov. Phys. JETP}\ }\textbf {\bibinfo {volume} {36}},\ \bibinfo {pages} {1193}
  (\bibinfo {year} {1973})}\BibitemShut {NoStop}%
\bibitem [{\citenamefont {Sun}(2010)}]{sun2}%
  \BibitemOpen
  \bibfield  {author} {\bibinfo {author} {\bibfnamefont {J.~J.}\ \bibnamefont
  {Su}}, \ and\ \bibinfo {author} {\bibfnamefont {A.~H.}\ \bibnamefont
  {MacDonald}}, \ }\href@noop {} {\bibfield  {journal} {\bibinfo  {journal} {Phys. Rev. B}\ }\textbf {\bibinfo {volume} {81}},\ \bibinfo {pages} {184523} (\bibinfo {year} {2010})}\BibitemShut {NoStop}%
\bibitem [{\citenamefont {Pesin}\ and\ \citenamefont
  {MacDonald}(2011)}]{pesin}%
  \BibitemOpen
  \bibfield  {author} {\bibinfo {author} {\bibfnamefont {D.~A.}\ \bibnamefont
  {Pesin}}, \ and\ \bibinfo {author} {\bibfnamefont {A.~H.}~\bibnamefont
  {MacDonald}},\ }\href@noop {} {\bibfield  {journal} {\bibinfo  {journal}
  {Phys. Rev. B}\ }\textbf {\bibinfo {volume} {84}},\ \bibinfo {pages} {075308}
  (\bibinfo {year} {2011})}\BibitemShut {NoStop}%
\end{thebibliography}
\end{document}